\documentclass[fleqn,usenatbib]{mnras}

\usepackage{mathptmx}

\usepackage[T1]{fontenc}

\DeclareRobustCommand{\VAN}[3]{#2}
\let\VANthebibliography\thebibliography
\def\thebibliography{\DeclareRobustCommand{\VAN}[3]{##3}\VANthebibliography}

\usepackage{graphicx}	
\usepackage{amsmath}	
\usepackage{amssymb}	
\usepackage{inputenc}
\usepackage{indentfirst}
\usepackage{comment}

\DeclareMathOperator*{\argmin}{arg\,min}

\title[On accretion in the eclipsing polar BS~Tri]{On accretion in the eclipsing polar BS~Tri}

\author[A. I. Kolbin et al.]{
Alexander I. Kolbin,$^{1}$\thanks{E-mail: kolbinalexander@mail.ru}
N. V. Borisov,$^{1}$
N. A. Serebriakova,$^{2}$
V. V. Shimansky,$^{2}$
\newauthor N. A. Katysheva,$^{3}$
M. M. Gabdeev,$^{1}$
S. Yu. Shugarov$^{4}$
\\
$^{1}$Special Astrophysical Observatory of Russian Academy of Sciences, Nizhnij Arkhyz, Karachai-Cherkessian rep., 369167 Russian Federation\\
$^{2}$Kazan (Volga-region) Federal University, Kremlevskaiya 18, Kazan, 420008 Russian Federation\\
$^{3}$Sternberg Astronomical Institute, Lomonosov Moscow State University, Moscow, 119991 Russian Federation\\
$^{4}$Astronomical Institute, Slovak Academy of Sciences, Tatranská Lomnica, 059 60, Slovak Republic 
}

\date{Accepted XXX. Received YYY; in original form ZZZ}

\pubyear{2020}

\begin{document}

\label{firstpage}
\pagerange{\pageref{firstpage}--\pageref{lastpage}}
\maketitle

\begin{abstract}
We analyze spectroscopic and photometric observations of the eclipsing polar BS~Tri. 
The polar's light curve shape variations can be interpreted by changing contributions
of the accretion stream to the integral radiation of the system. 
Based on the radial velocity curves of the irradiated part of the secondary, 
we refine the masses of the system components, $M_1 = 0.60 \pm 0.04 M_{\odot}$, $M_2 \approx 0.12 M_{\odot}$,
and the orbital inclination, $i=85\pm 0.5^{\circ}$.
The polar's spectra reveal cyclotron harmonics forming in an accretion spot 
with a magnetic field strength of $B=22.7 \pm 0.4$~MG and an average temperature of 
$T \sim 10$~keV. In addition to the cyclotron harmonics, the BS~Tri spectra contain 
Zeeman components of  H$\alpha$ line, which are probably formed in the cool halo near the accretion spot. 
The orientation of the magnetic dipole and the coordinates of the accretion spot are estimated 
by modeling the light curves of the polar. We show that for a satisfactory description of the 
BS~Tri light curves we have to take into account 
the variability of the spot's optical depth along the line of sight. 
Doppler maps of BS~Tri show a part of the accretion stream with a trajectory close 
to ballistic near the Lagrange point L$_1$, and another part of the stream 
moving along the magnetic field lines. 
The estimate of the stagnation region position found from the Doppler tomograms
is consistent with the photometric estimates of the accretion spot position.
 
\end{abstract}


\section{Introduction}

Polars (or AM~Her-type stars) are a type of cataclysmic variables characterized 
by a high magnetization of the white dwarf ($B\sim 10-200$~MG). 
Like other cataclysmic variables, polars are interacting close binaries, 
where the primary component is an accreting white dwarf, and the secondary 
is a cold (spectral type G-L) main-sequence star filling its Roche lobe. 
The matter of the secondary component outflows through the vicinity of the Lagrange 
point $L_1$ into the Roche lobe of the primary component and quickly reaches the stagnation region, 
where the dynamic pressure of the stream is compared with the magnetic pressure ($\rho v^2 = B^2/8\pi$).
The magnetic field prevents the motion of  ionized gas from the stream across the magnetic lines 
with the formation of an accretion disk. The accreted gas moves along the magnetic lines 
in the direction of the vicinity  of the white dwarf's  magnetic poles. See \citep{cropper90} for a review of AM~Her-type variables.

The gas falling at a supersonic velocity forms a shock front near the surface of the star, 
where the matter is heated to high temperatures  ($T=10-50$~keV). 
After passing through the shock, the gas  settles on the surface of the white dwarf at a subsonic velocity, 
cooling by means of bremsstrahlung X-ray radiation and optical cyclotron radiation. 
The height of the shock front above the surface of the star is determined by the 
accretion rate and the magnetic field strength. Usually it is  
$\sim 0.01-0.1$ of the radius of a white dwarf. Due to the density inhomogeneity of the stream, 
its transition region of gas to the magnetic trajectory is stretched, 
resulting in elongation of spots along the surface of the star \citep{mukai88}.


BS~Tri was discovered as a bright X-ray source by the ROSAT space observatory. It 
is cataloged as object 1RXS J020928.9+283243. Spectroscopic observations of BS~Tri 
were carried out by \cite{wu01}. The obtained spectrum revealed 
the Balmer series hydrogen and neutral helium emission lines 
typical of cataclysmic variables, as well as the line of ionized helium 
HeII $\lambda$~4686. Photometric observations of BS~Tri were carried out at the 1.5-m 
Russian-Turkish telescope by  \citet{denis06} within the program for 
studying the X-ray sources from the ROSAT catalog. 
The resulting light curve of BS~Tri contained deep  ($\Delta V \approx 4.5^m$) 
eclipses lasting about $\Delta t_{ecl} \approx 6$~min. 
In the same study, the orbital period $P_{orb} = 96.26$~min was determined. 
The magnetic nature of BS~Tri was first suggested by \cite{rodriguez05}, who 
interpreted the two-peak structure of the light curve as a feature of cyclotron 
radiation generated in an accretion spot. Spectroscopic studies of BS~Tri were carried out 
by \cite{borisov15}. The authors have shown that emission lines are formed both in the accretion stream
and on the irradiated hemisphere of the red dwarf due to the reprocessing effects. 
Based on the red dwarf's irradiated hemisphere radial velocities, 
the masses of the system components were estimated as
$M_1 = 0.75 \pm 0.02M_{\odot}$, $M_2 = 0.16 \pm 0.01 M_{\odot}$.

In this study, we analyze the spectral and photometric manifestations of accretion in BS~Tri. 
The second section of our paper describes the spectroscopic and photometric observations 
of BS~Tri, as well as the methods of processing the obtained material. 
The morphological analysis of the light curves of the polar is given in Section \ref{morph}. 
Then, in Section \ref{syspars}, the masses of the system components and 
orbital parameters are refined. The analysis of the behavior of cyclotron harmonics 
and determination of  magnetic field strength of the white dwarf are laid out 
in Section \ref{parcycspec}. Section \ref{parlcmod} describes  the  technique
we propose for modeling the polar curves and the results of its application to BS~Tri. 
The results of Doppler tomography of the polar are discussed in Section \ref{pardoptom}. 
The results of the study itself are summarized in the conclusion.

\section{Observations and data reduction}

{\bf Spectroscopy.} A set of BS~Tri spectra was obtained with the 6-m BTA telescope 
of the Special Astrophysical Observatory of the Russian Academy of Sciences 
on the nights of 2011 September 21/22 (18 spectra) and 2012 August 26/27 (31 spectra) using the SCORPIO-2 
focal reducer in the long slit spectroscopy mode \citep{afan}. 
In both sets of observations, a VPHG1200@540 holographic diffraction grating was used.
However, in the first set, the slit size was $1''$, providing a spectral resolution of 
$\Delta \lambda = 5.2$~\AA/pixel, while in the second set, the slit was narrowed to 
$0.5''$ with the corresponding resolution of $\Delta \lambda = 2.6$~\AA/pixel. 
The spectra were recorded on an E2V CCD42-90 nitrogen-cooled CCD ($2088\times 4632$ pixels), 
covering the spectral range of $\lambda \lambda = 3700-7300$~\AA~with  a
$2\times 1$ binning. The observations were carried out in good astroclimatic conditions
with an average seeing of $1.0''$ and $1.2''$ for 2011 and 2012, respectively. 
All BS~Tri spectra were obtained with an exposure of 300~s.

The reduction of spectral data was carried out in the IRAF environment
\footnote{The IRAF astronomical data processing and analysis package was developed 
by the National Optical Astronomical Observatory, USA. The package is available at: 
https://iraf-community.github.io.} and included all the standard procedures for the reduction 
of spectroscopic observations. The traces of cosmic rays were removed from the images 
using the LaCosmic algorithm\footnote{LaCosmic programs for the removal of cosmic ray traces
from the  astronomical images  are available at: http://www.astro.yale.edu/dokkum/lacosmic.} 
\citep{dokkum01}, the electron bias was subtracted. Based on the flat-field lamp images, 
the inhomogeneity of the CCD sensitivity was reduced, and the geometric distortions 
introduced by the spectrograph optics were corrected. Optimal extraction of the object 
spectra with the background subtraction and their wavelength calibration based on 
He-Ne-Ar lamp frames have been performed. 
Spectrophotometric calibration was performed based on the images of the standard star. 
For each spectrum, barycentric Julian dates and barycentric corrections to  
radial velocity were calculated.

\begin{figure}
\center
\includegraphics[width=\columnwidth]{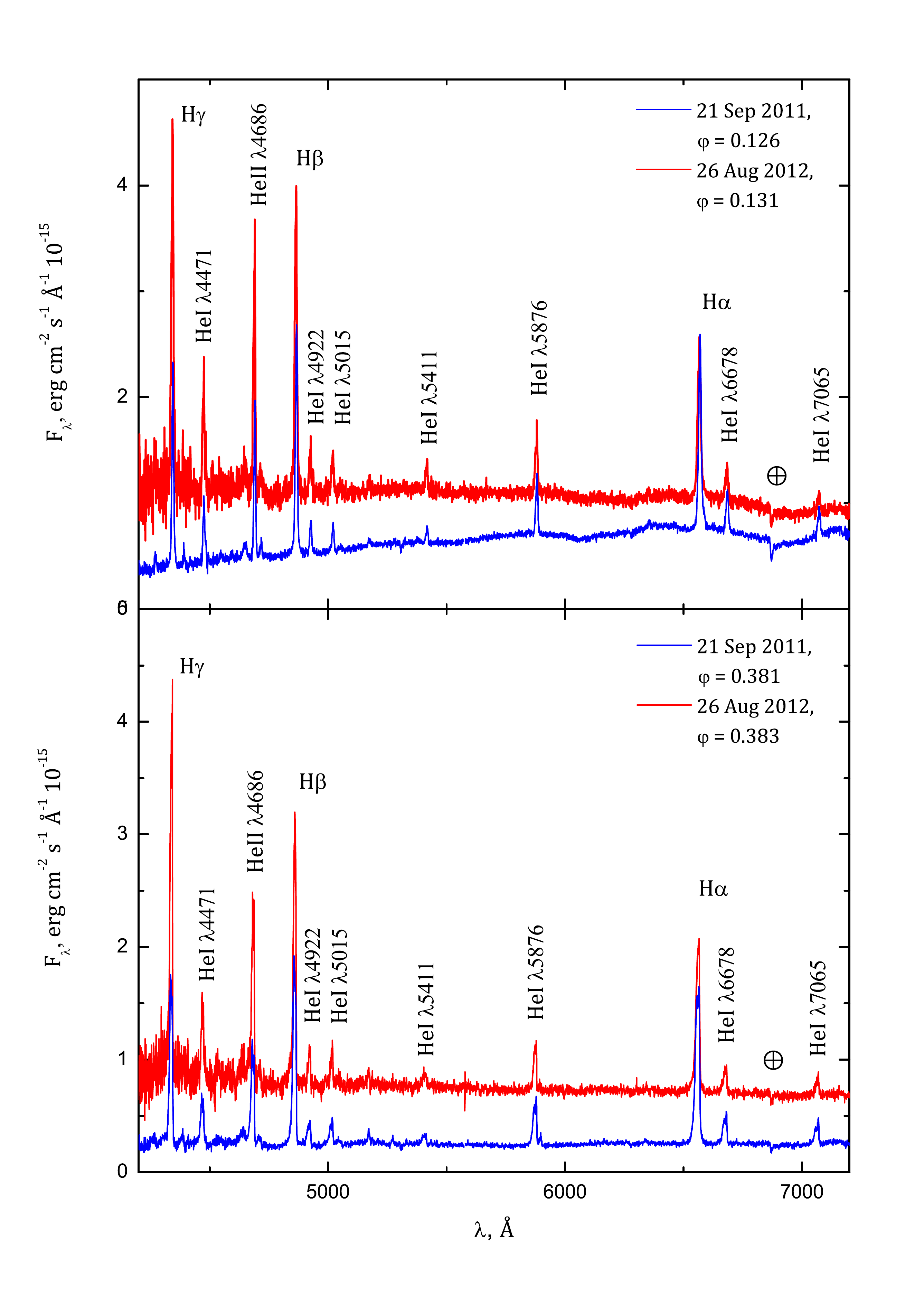}
\caption{Examples of BS~Tri spectra obtained in the 2011 and 2012 observations. 
The upper panel shows spectra obtained at phases close to the maximum brightness ($\varphi=0.126$ and $\varphi=0.131$). 
The bottom panel shows the spectra obtained near the minimum brightness of the object 
(phases  $\varphi=0.381$  and $\varphi=0.383$). The 2012 spectra have been shifted upwards by $0.5 \times 10^{-15}$ erg cm$^{-2}$ s$^{-1}$ \AA$^{-1}$. }
\label{specs}
\end{figure}

The obtained spectra have shapes typical of the AM~Her-type variables, 
including the hydrogen emission lines in the Balmer series, the lines of neutral and 
ionized helium. Examples of the spectra obtained in two sets of observations 
in the phases of maximum and minimum brightness are shown in Fig \ref{specs}. 
A change of slope is noticeable for spectra obtained during the observations in 2011 and 2012 near the phase of brightness maximum. In the 2011 observations, there were a   
 red-ward slope and broad ``humps'', interpreted as cyclotron harmonics. 
 In 2012, the slope changed and the maximum shifted to the blue region. The cyclotron harmonics became less pronounced. 
 No changes are observed in the slope of spectra near the 
 phase of the minimum.

{\bf Photometry.} Photometric data for BS~Tri was obtained with the ZTE 125-cm telescope 
and the Zeiss-600 60-cm telescope, which are located at the South Astronomical Station 
of Moscow State University. Additional observations were carried out with the Zeiss-1000 
1-m telescope of the Special Astrophysical Observatory of the Russian Academy of Sciences. 
BS~Tri aperture photometry was performed using the Maxim~DL software. 
The photometric observations log is presented in Table~\ref{log_phot}.

\begin{table}
\caption{Log of photometric observations of BS~Tri. (L --- observations without a filter).}
\label{log_phot}
\begin{center}
\begin{tabular}{lllc}
\hline
Telescope 	& Date (UT)		& JD 2450000+				& Filter 	\\ \hline
ZTE			& 2005 Nov 26	& 3701.2799 -- 3701.4062 	& B \\
ZTE			& 2005 Nov 26	& 3701.2824 -- 3701.3788 	& V \\
ZTE			& 2005 Nov 26	& 3701.2799 -- 3701.4062	& R \\
ZTE			& 2005 Nov 26	& 3701.2789 -- 3701.3777 	& I \\
Zeiss-600	& 2011 Nov 26	& 5831.4536 -- 5831.5170	& V \\
Zeiss-600	& 2011 Dec 02	& 5898.3927 -- 5898.5123	& V \\
ZTE 		& 2018 Nov 09	& 8432.2736 -- 8432.3793 	& L \\
ZTE			& 2018 Nov 10	& 8433.2376 -- 8433.3072 	& L \\
ZTE			& 2018 Nov 11	& 8434.2481 -- 8434.3795 	& L \\
ZTE			& 2018 Nov 14	& 8434.3795 -- 8437.3663 	& L \\
Zeiss-1000	& 2019 Aug 29-31	& 8725.4230 -- 8727.5415 	& Rc \\
Zeiss-1000	& 2019 Aug 29-31	& 8725.4211 -- 8727.5437 	& V \\
\hline
\end{tabular}
\end{center}
\end{table}

\section{Photometry analysis}
\label{morph}

The obtained light curves were converted from the Julian date scale to the orbital phase scale
using the period $P_{orb} = 0.0668810424(2)$~days, obtained by \cite{borisov15}. 
Unfortunately, the ephemerides presented in the same work could not successfully predict the 
time of eclipses in BS~Tri. For this reason, the initial epoch was selected individually 
for each set of observations. Note that it was also not possible to calculate new 
ephemerides over a long time interval due to unidentified problems with the 
time support at the ZTE.

\begin{figure}
\center
\includegraphics[width=\columnwidth]{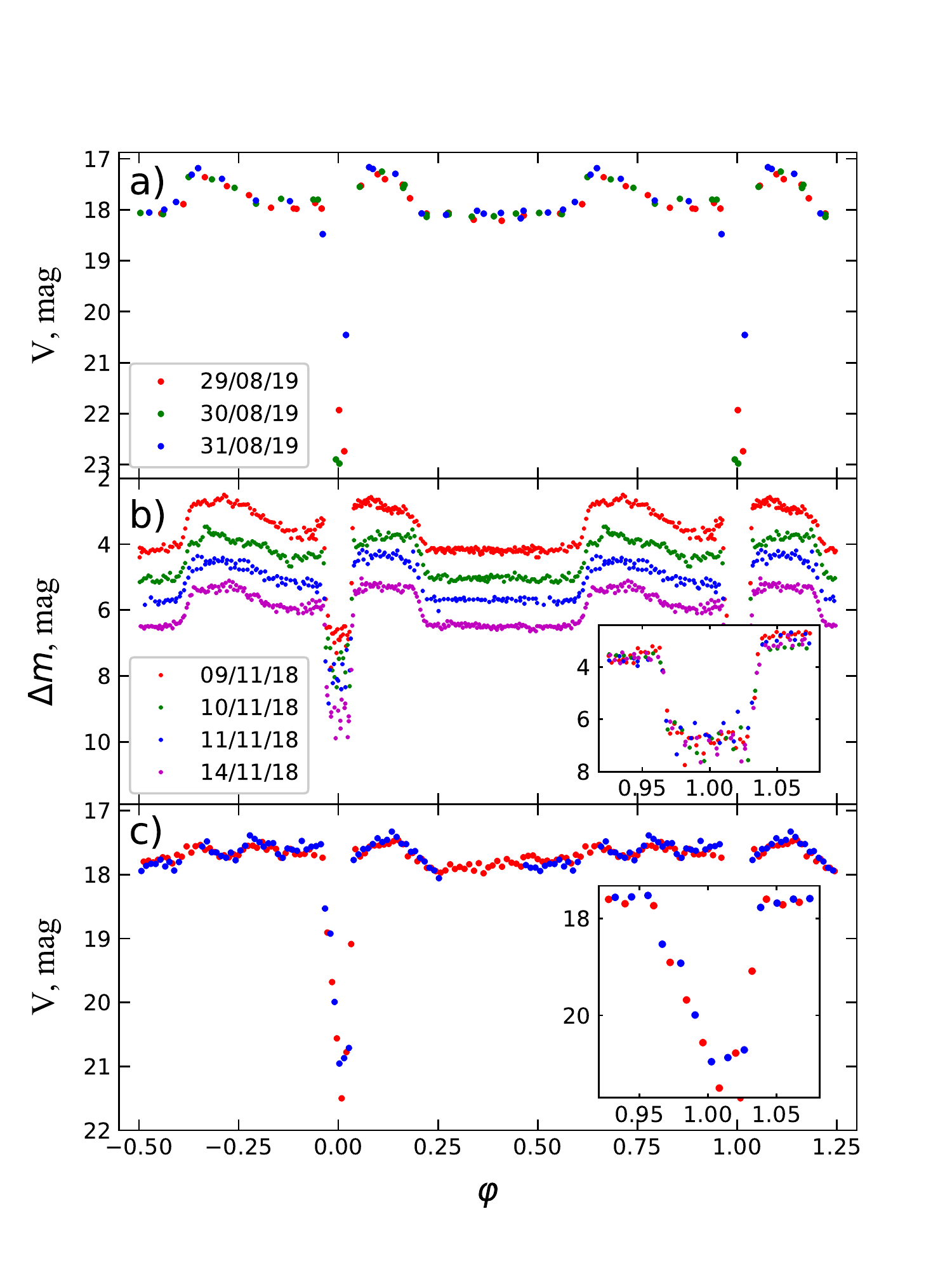}
\caption{Examples of light curves of BS~Tri. a) Light curves obtained with the Zeiss-1000 telescope 
on 2019 August 29-31 (lowered state). b) Light curves of the polar obtained with the ZTE telescope on 2018 November 9-14 (lowered state) without photometric filters. The data for adjacent nights are offset from each other by
$0.75^m$. The subplot shows the polar eclipse profile on a larger scale. 
c) The light curve obtained from the observations with the Zeiss-600 telescope on 2011 December 2 (heightened state). 
The subplot shows the  profile of the eclipse in the polar.}
\label{lcs}
\end{figure}

Two extreme cases can be distinguished by morphology in the available BS~Tri light curves, 
which we shall call the ``lowered'' and ``heightened'' states.  
Note that we do not use the terms low and high states, implying a significant 
(by several stellar magnitudes) change in the average brightness of the polar. 
In our case, no strong variations in the average brightness were recorded 
for the entire observation period of the polar. In the lowered state, 
the light curve has a rather simple shape and is divided into two parts: 
a bright double-humped maximum, covering the phase interval $\varphi=0.60-1.21$, 
and a horizontal plateau extending over $\Delta \varphi \approx 0.4$. 
An example of such a light curve obtained with the Zeiss-1000 telescope in the $V$-band is 
shown in Fig.~\ref{lcs}~a. The available light curves of the lowered state 
in the $V$-band unfortunately have a low temporal resolution for analyzing the eclipse profile. 
Therefore, in Fig.~\ref{lcs}~b we present similar light curves obtained with the
ZTE without filters, but with higher resolution. 
We can see that the eclipse profile is symmetric. In the heightened state, the light curve 
has a more complex structure, the extended plateau is absent, the bright phase is 
less pronounced and has a complex asymmetric shape. An example of such a light 
curve obtained with the Zeiss-600 is shown in Fig.~\ref{lcs}~c. The eclipse profile 
has a smooth ingress and a quick egress. Although there are significant differences in the shapes of the light curves, the brightness of the polar does not change much between the two states. Thus, in the phase range $\varphi=0.25-0.50$, the brightness in the heightened state is $\sim 0.2$~mag higher than in the lowered one. 

We may assume from the above that in the lowered state the main source of out-of-eclipse 
variability is the accretion spot, whose radiation is of cyclotron nature. 
It passes across the disk of the white dwarf during its maximum brightness ($\varphi=0.60-1.21$). 
The intensity of cyclotron radiation is maximal  when radiation is perpendicular to 
the magnetic field lines, which explains the two-humped structure of the bright phase. 
If the magnetic field lines are normal to the surface of the white dwarf, 
then the brightness maxima should be observed near the moments of entry into and exit from the visible disk 
of the star, which is observed in the lowered state of BS~Tri. When the spot is behind the limb of the star, 
the main source of radiation is the white dwarf. The secondary component and the weak stream do not
yield significant brightness variability, which explains the presence of a wide plateau. 
In the heightened state, the accretion stream makes a noticeable contribution to the polar emission. 
Its coverage by the secondary component explains the smooth entry into the eclipse light curve. 
Due to a large optical thickness of the stream, it provides brightness variability during the 
entire orbital period, which explains the absence of a plateau in the  increased state. 
In addition, in this state there are indications of the existence of a weak dip ($\varphi=0.86$), which is formed due to the covering of the accretion spot by the stream.

The eclipse duration $\Delta t_ {ecl}$ was determined based on the light curves obtained with the ZTE 
telescope without filters. To determine this duration, we described the brightness minimum
by an inverted trapezoid
with lateral sides symmetric to the center of the eclipse. The upper part of the trapezoid 
was cut off by a straight line describing the adjacent out-of-eclipse parts of the light curve. 
The parameters of the trapezoid were found by the least squares method. 
The found eclipse duration (i.e. the width of eclipse profile at half depth) $\Delta t_{ecl} = 415 \pm 7$~sec, 
its depth $\Delta m = 3^m.48 \pm 0^m.15$, and the upper limit of the duration of 
the eclipse ingress and egress is $\Delta t = 42\pm 14$~sec.  

\section{BS~Tri parametres}
\label{syspars}

\begin{figure}
\center
\includegraphics[width=\columnwidth]{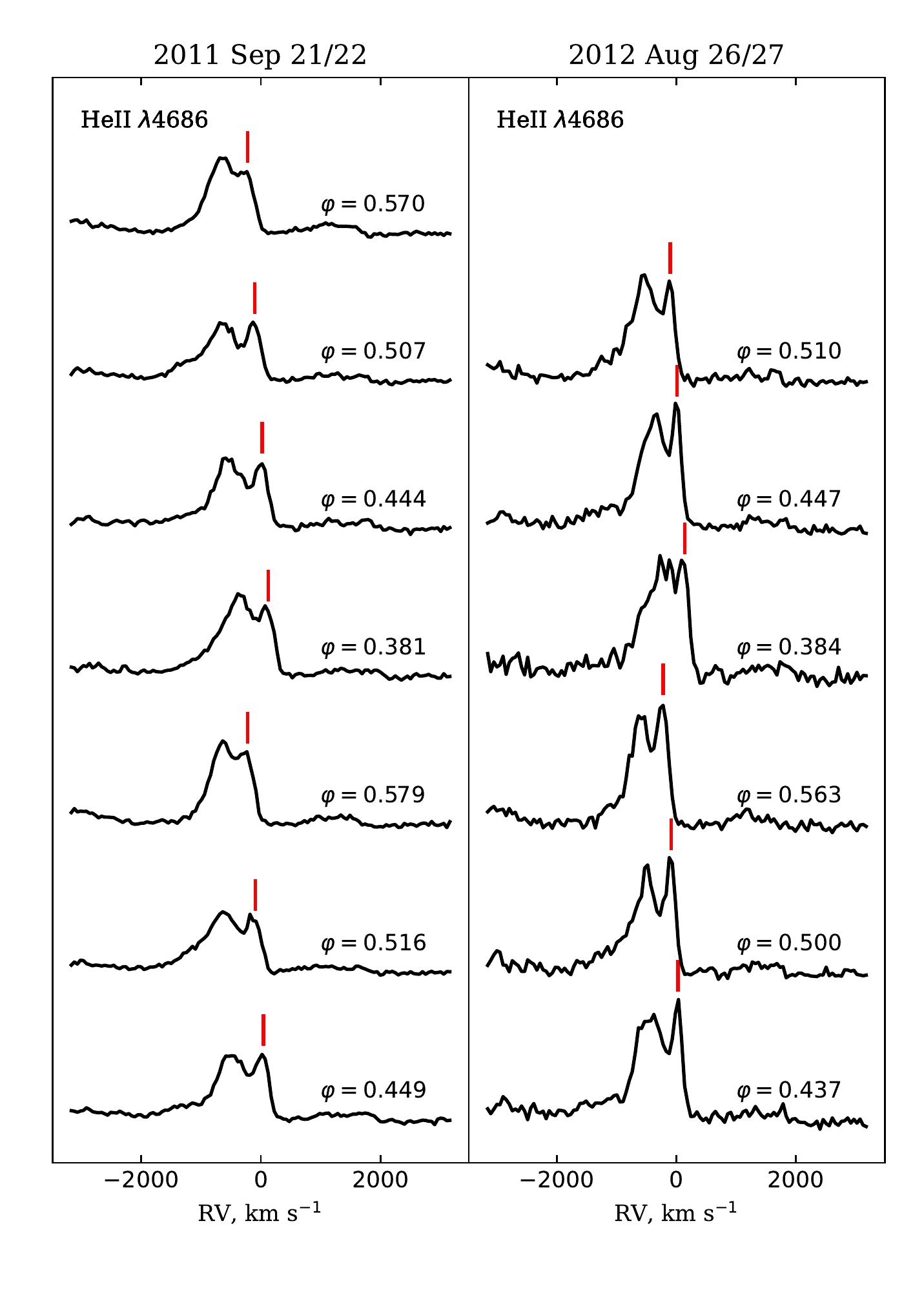}
\caption{The line profiles of HeII $\lambda$4686 containing narrow emission component. The positions of narrow emission componnent are marked by vertical lines.}
\label{nel}
\end{figure}

\begin{figure}
\center
\includegraphics[width=\columnwidth]{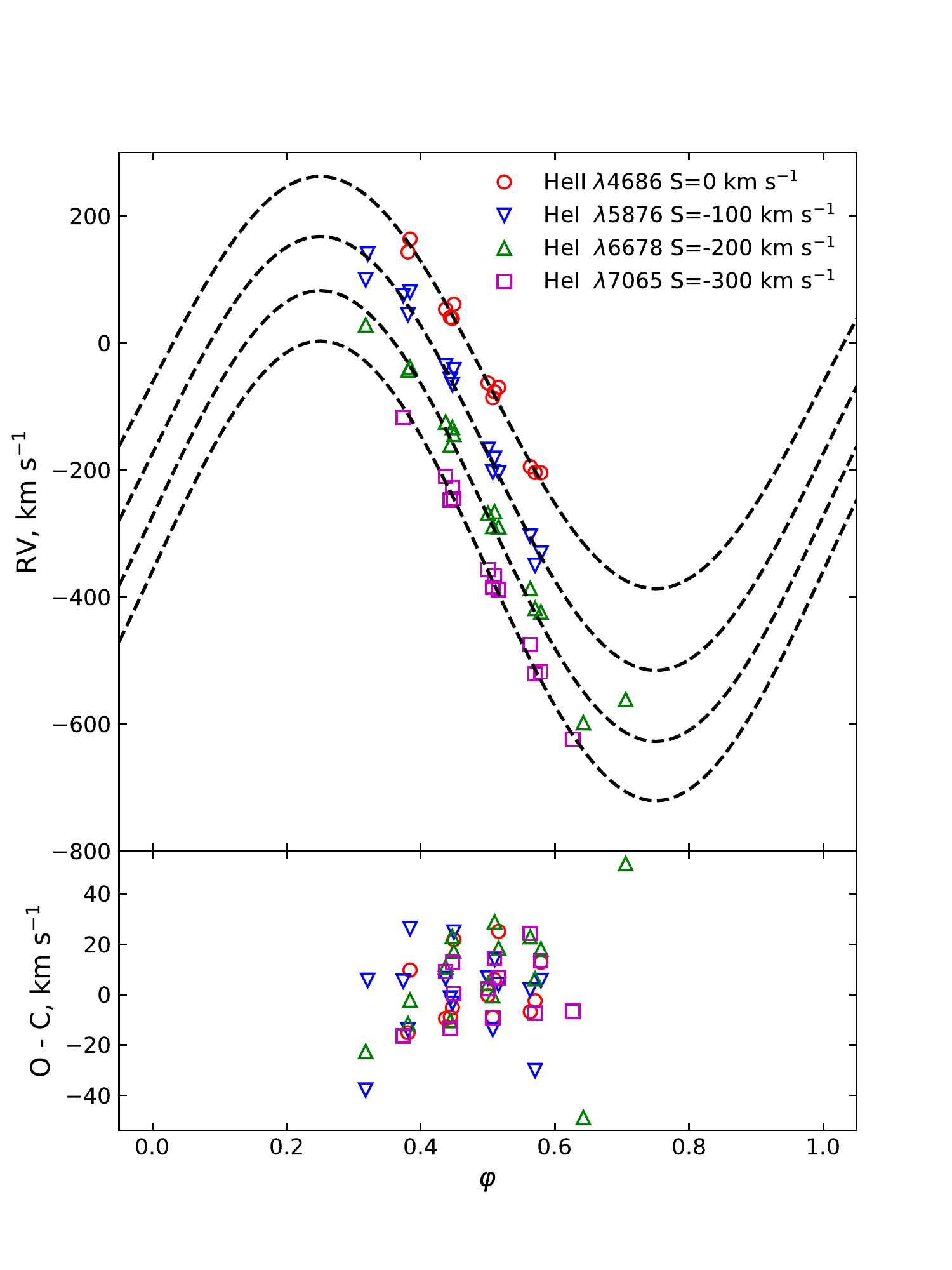}
\caption{Top panel: radial-velocity curves of narrow components of neutral and ionized helium lines 
(for clarity, the curves are shifted vertically by $S$); the dotted line describes the 
sine wave approximation. 
Bottom panel: radial velocity deviations of narrow line components from the approximating sinusoid.}

\label{rvs}
\end{figure}

In the range of the orbital phases $\varphi=0.30-0.65$, narrow emission components are observed in the 
lines of hydrogen and helium, moving from the red to the blue region. These components are formed on the 
surface of the red dwarf, heated by X-ray radiation from the accretion spot \citep{borisov15}. 
They are most pronounced in the lines of neutral (HeI~$\lambda$5876, HeI~$\lambda$6678, HeI~$\lambda$7056) 
and ionized (HeII~$\lambda$4686) helium. As an example, the Fig. \ref{nel} shows HeII~$\lambda$4686 line profiles with a pronounced narrow emission component.  The radial velosities of narrow components were measured by approximating the line profiles 
with a set of Gaussians. 
The measurements were carried out using the data from two sets of spectral observations, 
and the profiles with poorly separated components were excluded from analysis. The average error of radial velosity determination of narrow emission component is $\sim 10$~km/s for HeII~$\lambda4686$ and $\sim15$~km/s for the lines of neutral helium. The radial-velocity curves obtained 
this way are shown in Fig.\ref{rvs}.

The observed radial velocities $V_r$ of the narrow component of the emission lines were approximated by the circular orbit model:

\begin{equation}
V_r(\varphi) = \gamma + K'_2 \sin(2\pi \varphi),
\end{equation}
where $K'_2$ is the radial velocity semi-amplitude of the red dwarf's irradiated hemisphere, $\gamma$ is the 
radial velocity of the system's center of mass, $\varphi$ is the phase of the orbital period. 
Varying the initial phase $\varphi_0$ when approximating the radial velocities did not lead 
to a significant variation in $\gamma$ and $K'_2$, which is why it was considered
to be $\varphi_0 = 0$.

The radial velocities of the center of mass of the system $\gamma$, as well as the radial velocity
semi-amplitudes of the narrow components of the $K'_2$ lines, found by the least-squares method 
are presented in Table~\ref{lines}. A comparison of the approximating sinusoids with the measured 
radial velocities is shown in Fig.~\ref{rvs}. The errors in determining $K'$ and $\gamma$ were 
calculated using the  
theory of back-propagation of errors for linear inverse problems (see, e.g., \cite{aster}).

To determine the masses of the system components, it is important to find the 
radial velocity semi-amplitude of the red dwarf's center of mass $K_2$. 
The $K_2$ semi-amplitude was 
found by the modeling of the X-ray irradiated atmosphere in different regions of the red dwarf's surface according to the technique of \cite{shim12}. The rotation rate of the red dwarf was assumed to be synchronous with the orbital motion and the parameters of the atmosphere were taken as $ T_{eff} = 3250 $~K, $ \log g = 5.0 $. The X-ray luminosity of the accretion spot was fixed at $L_{X} = 10^{32}$ erg s$^{-1}$ as the most probable for polars \citep{barrett99}, and its energy distribution was taken according to the data of \cite{cropper90} for AM~Her systems. The theoretical radial velocities of the studied HeI and HeII lines were measured by the cross-correlation technique for 72 phases of the orbital period and fitted by a sinusoid for the determination $\Delta K_2 = K_2 - K'_2$ corrections for each line. The corrections $\Delta K_2$ found for each analyzed helium line are listed in Table \ref{lines}. 
The average value of the radial velocity semi-amplitude of the center of mass, 
found from the lines of neutral and ionized helium, is $K_2 = 390 \pm 11$ km/s. 
The mass function corresponding to the found $K_2$  is $f(m) = 0.411 \pm 0.003 M_{\odot}$.

\begin{table}
\caption{Approximation parameters of the radial-velocity curves of narrow components of spectral lines, 
as well as corrections for the reduction to the center of mass $\Delta K_2$.}
\label{lines}
\begin{center}
\begin{tabular}{lccc}
\hline
Line 				& $\gamma$, km s$^{-1}$ 	& $K'_2$, km s$^{-1}$	& $\Delta K_2$, km s$^{-1}$	\\ \hline
HeII~$\lambda$4686	&$-63 \pm 4$				& $325 \pm 9$			& 50.3 	\\
HeI~$\lambda$5876 	&$-74 \pm 5$				& $342 \pm 9$			& 42.7 	\\
HeI~$\lambda$6678	&$-73 \pm 6$				& $355 \pm 11$			& 43.3 	\\
HeI~$\lambda$7056	&$-59 \pm 3$				& $362 \pm 8$			& 41.4 	\\
\hline
\end{tabular}
\end{center}
\end{table}

Additional constraints necessary for determining the parameters of the BS~Tri components 
can be obtained from the fact that the red dwarf fills its Roche lobe. According to \cite{eggleton83}, 
the effective radius $R_L$ of the star filling the Roche lobe can be estimated using the formula
 
\begin{equation}
\frac{R_L}{A} = 0.49 \frac{Q^{-2/3}}{0.6Q^{-2/3}+\ln (1+Q^{-1/3})},
\label{eqql}
\end{equation}
where $Q=M_1/ M_2$, i.e. the ratio of masses of the primary and the secondary, 
and $A$ is the separation
of the centers of mass of the components determined with Kepler's third law: $A = (M_2(Q+1)P^2_{orb})^{1/3}$.

On the other hand, the radius of the secondary located on the main sequence is related to its mass. 
To plot the ``Radius--Mass'' $R'(M_2)$ dependence, we used the evolutionary tracks of \cite{girardi00} 
for the solar abundance of elements ($Z=0.019$) and taking into account convective overshooting. 
By solving the equation $R'(M_2) = R_L (Q, M_2)$  we can find a set of solutions in the plane 
($Q$--$M_2$). The secondary components located in the cataclysmic systems have high rotation velocities 
due to the tidal synchronization, as well as deep convective shells. Both of these factors lead to 
their high magnetic activity due to the magnetic dynamo mechanism \citep{cropper90}. 
Such stars should have high photospheric spottedness, which prevents the convective transfer of energy 
in the atmosphere, and, as a consequence, an increased radius compared to ``normal'' stars for the 
transfer of the energy flux generated in the core. To take this effect into account, we multiplied the radii 
$R'(M_2)$ by the factor of $k=1.15$ found from the observed ``Radius--Period'' dependence of the rapidly rotating 
($P\le 1$~days) dwarfs from the Pleiades \citep{somers17}.

\begin{figure}
\center
\includegraphics[width=\columnwidth]{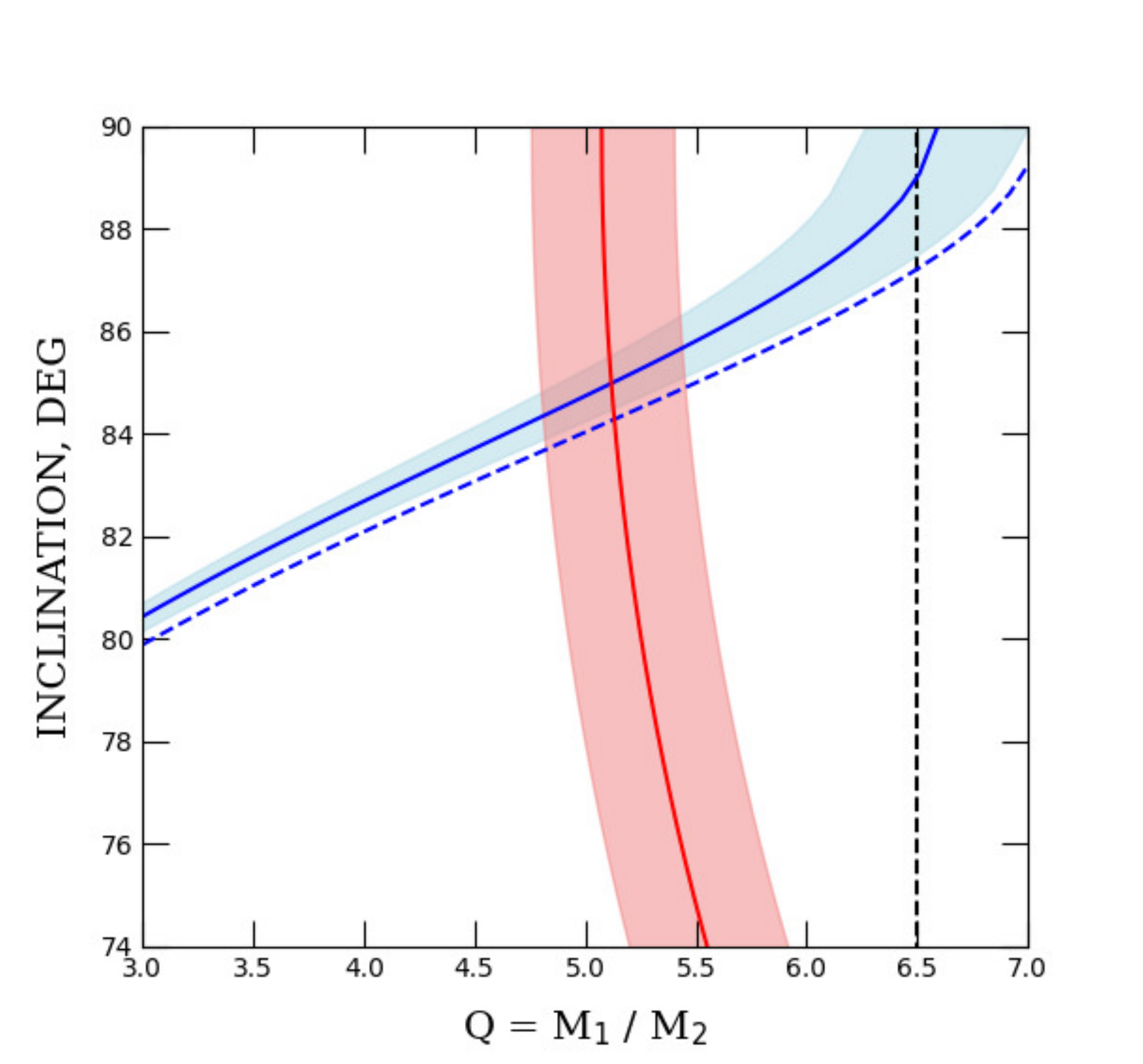}
\caption{Solution sets in the plane ($Q$--$i$) found from the duration of the eclipse (the blue curve) and 
from the radial-velocity curve of the narrow component HeII~$\lambda$4686 (the red line). 
The blue and red areas limit the confidence regions of the desired parameters. 
The blue dashed line corresponds to the solution providing the observed duration of the eclipse for 
the case of the maximum proximity of the accretion spot to the secondary component. 
The vertical dashed line shows the solution obtained by describing the ballistic part of the stream on the Doppler tomogram.}
\label{qi}
\end{figure}

For the convenience of further reasoning, we have constructed  in Fig.~\ref{qi} a set of solutions 
in the plane ($Q$--$i$), where $i$ is the angle of inclination of the orbital plane to the line of sight. 
This set is obtained by equating the observed mass function $f(m)$ to the theoretical one:
\begin{equation}
f(m) = \frac{M_2 Q^3 \sin^3 i}{(Q+1)^2},
\end{equation}
where the mass $M_2$ was determined by solving the equation $R'(M_2) = k R_L(Q, M_2)$ at fixed $Q$ values. 
In addition, the figure shows the confidence region constructed from the errors in determining the 
radial velocity semi-amplitude of the narrow component of $K'_2$ lines.

Figure~\ref{qi} shows a set of solutions in the plane ($Q$--$i$), which provides the observed duration 
of the eclipse $\Delta t_{ecl} = 415 \pm 7$~sec in BS~Tri. 
To calculate this set, the surface of the secondary was assumed to repeat the surface of the Roche lobe. 
The phases of entry and exit from the eclipse were determined from the requirement that the red dwarf's Roche 
lobe touches the line of sight passing through the centre of the white dwarf. 
A significant source of errors in determining the angle $i$ may be the assumption that 
the white dwarf's radiation dominates in the optical range. In fact, a bright accretion spot 
contributes to the observed duration of the eclipse. In order to estimate the polar parameter 
determination error, associated with the uncertainty in the position of the main radiation source, 
we placed a point source of radiation on the white dwarf surface facing the secondary component. 
The solution obtained for the eclipse of this source in the plane ($Q$--$i$) is shown in Fig.\ref{qi} 
with a dashed line. We can see that the corresponding uncertainty in the inclination angle is about
$\Delta i \approx 0.75^{\circ}$.

The intersection of two curves in the ($Q$--$i$) plane occurs at a point with the coordinates 
$Q = 5.12 \pm 0.35$  and $i=85\pm0.5^{\circ}$.
The agreement between the size of the Roche lobe and the evolutionary radius of the secondary is 
achieved at the mass $M_2 = 0.116$ (the error in $Q$ does not lead to a significant uncertainty in $M_2$). 
The radius of the star corresponding to a given mass is $R_2 = 0.155R_{\odot}$.
From the found $Q$ and $M_2$ stems the mass of the white dwarf $M_1 = 0.60 \pm 0.04 M_{\odot}$, 
while its radius, according to the ``Radius--Mass'' relation \cite{nauenberg72}, is equal to 
$R_1 = 0.0125 \pm 0.0005 R_{\odot}$. The distribution of the component centers of mass 
obtained according to Kepler's third law is equal to $A = 0.62 \pm 0.01 R_{\odot}$.

\begin{figure}
\center
\includegraphics[scale=0.5]{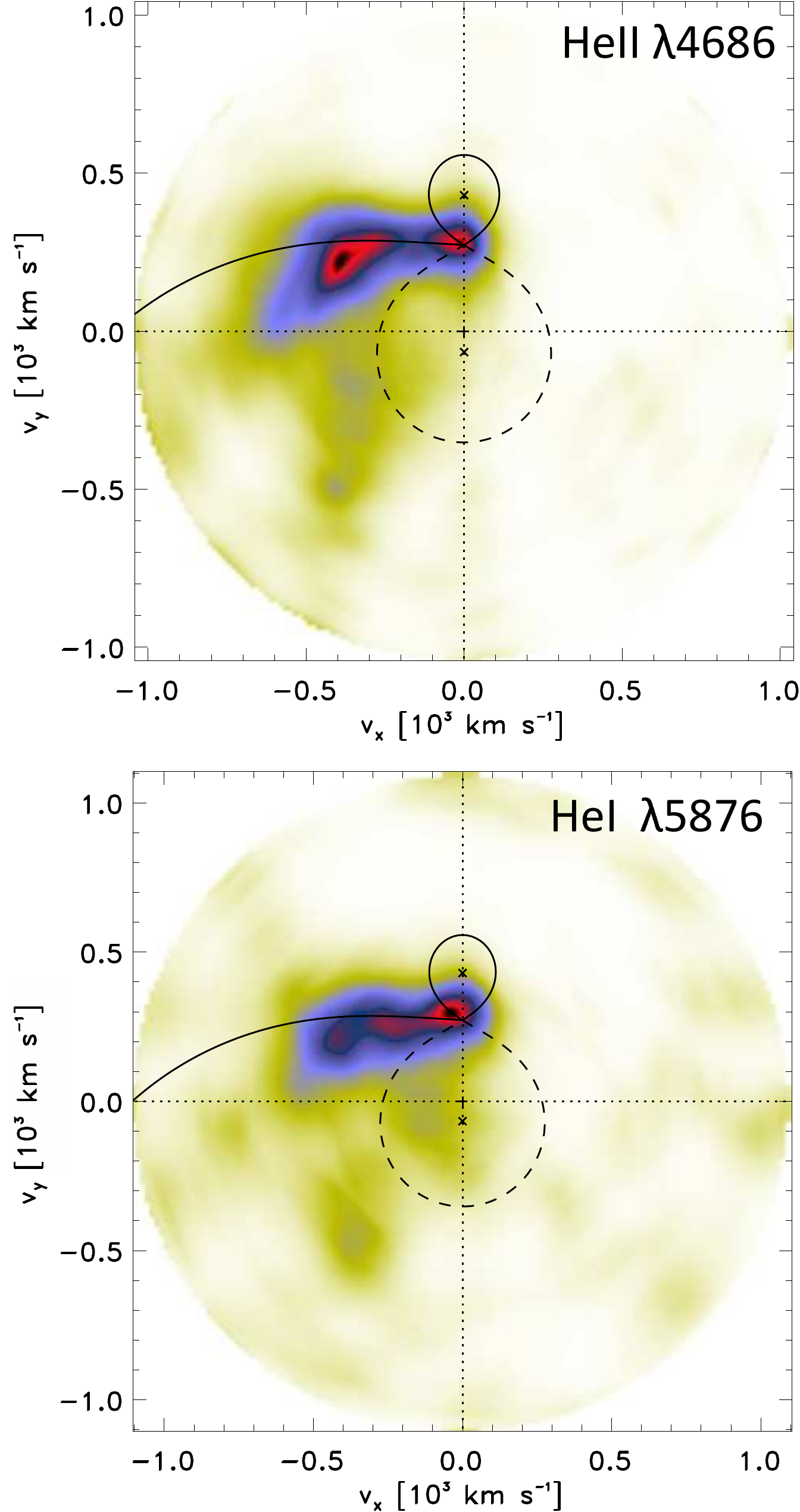}
\caption{Doppler tomograms of BS~Tri in the HeII~$\lambda$4686 and HeI~$\lambda$5876 lines 
based on the 2012 spectroscopic observations. The tomograms are superimposed with the traces of the 
Roche lobes of the primary (the dashed closed curve) and secondary (the continuous closed curve) components of the system, 
as well as the trace of the theoretic ballistic trajectory (the continuous line) for the mass ratio $Q=6.5$. 
The crosses indicate the centre-of-mass velocities of the system components.
}
\label{dopptom1}
\end{figure}

One of the methods for estimating the masses of the  cataclysmic system components
is based on the analysis of Doppler tomograms, which reveal  the ballistic part 
of the accretion stream. 
Apparently, this part manifests itself on the BS~Tri Doppler tomograms in the HeII~$\lambda$4686  and HeI~$\lambda$5186 
lines, obtained from the   2012 spectroscopic observation data. These tomograms are shown in Fig.\ref{dopptom1}. 
The Doppler tomography method applied to BS~Tri is described in details in Section \ref{pardoptom} of this work, 
while for the details of tomogram interpretation we refer the reader to the works of \citep{marsh88, kotze15}. 
The parameters of the system were estimated by fitting the theoretical velocities of the ballistic trajectory 
to the stream section lying in the velocity range of
$-400$~km/s $\le v_x\le 0$~km/s. For $v_x < -400$~km/s, a strong deviation of the trajectory towards smaller 
$v_y$ is observed, which is apparently associated with reaching the Alfv\'en radius and the onset 
of gas motion along the magnetic field lines. The ballistic trajectory is determined by the ratio of masses $Q$ 
and the mass of one of the components, for example, $M_2$. To eliminate the ambiguity in determining the 
parameters, the mass of $M_2$ was determined by matching the size of the Roche lobe with the evolutionary 
radius of the star at a given $Q$: $R'(M_2) = k R_L(Q, M_2)$. 
The best agreement of the theoretical stream velocities with the observed ones was achieved for $Q \approx 6.5$. 
This estimate contrasts strongly with the $Q$ value found above. It is possible that the reason for this 
discrepancy is associated with a more complex behavior of the stream near the vicinity of the Lagrange point L$_1$, 
which was noted by \cite{schwarz02, staude04, schwope00} when analyzing the Doppler tomograms of AM~Her and QQ~Vul.

\section{Cyclotron spectra modeling}
\label{parcycspec}

\begin{figure}
\center
\includegraphics[width=\columnwidth]{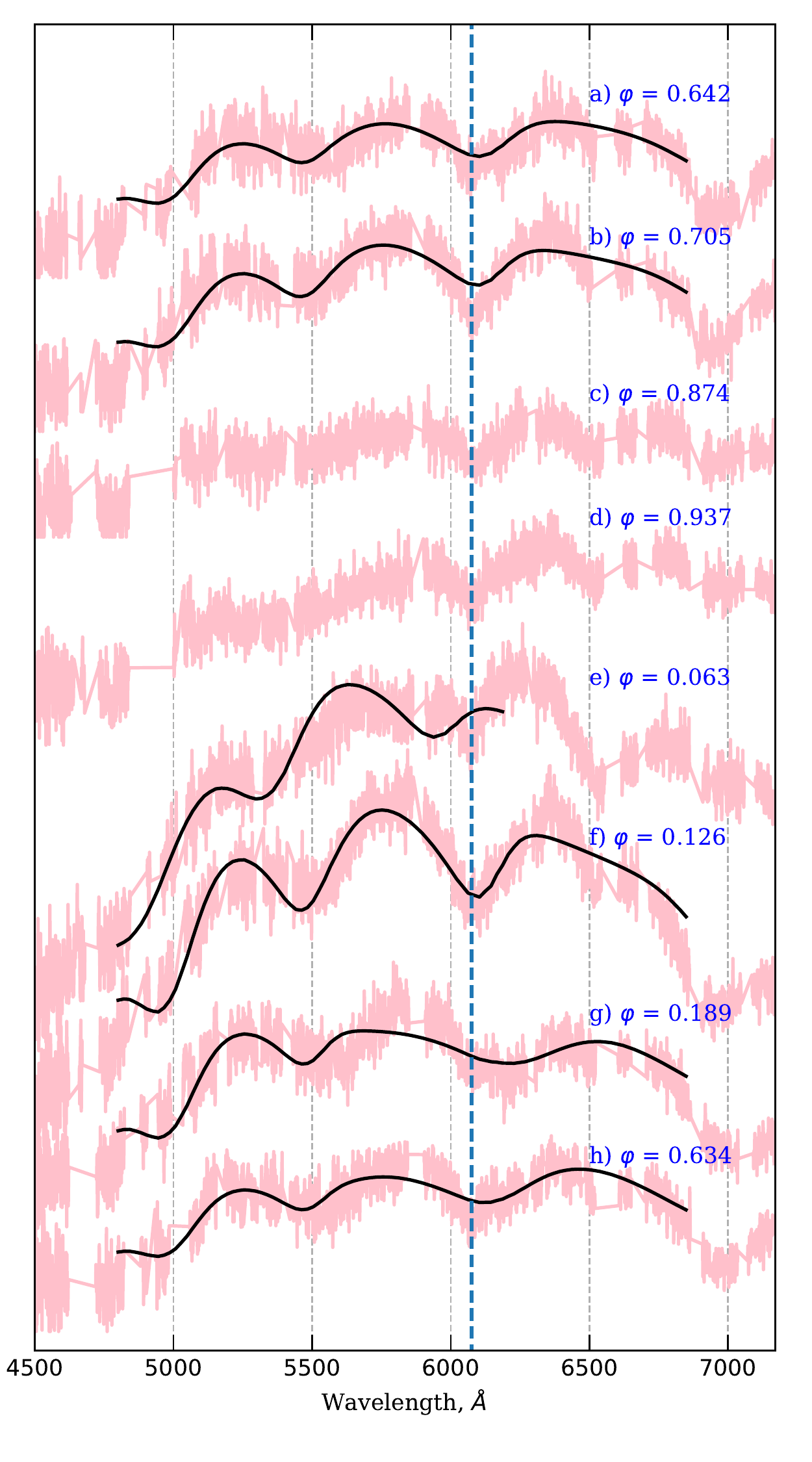}
\caption{The observed BS~Tri spectra (pink lines) with a subtracted non-cyclotron component formed outside the accretion spot, as well as their fit by the theoretical cyclotron spectra (black lines). The position of $6075$~\AA~absorption feature is marked by a vertical dashed line.}
\label{harms}
\end{figure}

As mentioned earlier, the BS~Tri spectra of the 2011th year contain broad ``humps'' that can be interpreted 
as cyclotron harmonics. These features appear in the range of the orbital phases $\varphi=0.63-0.19$ 
corresponding to the transit of the accretion spot across the observed disk of the white dwarf. 
The spectra of the specified phase range are shown in Fig. \ref{harms}. 
The regions of emission lines were removed, and the continuum of the average spectrum at the brightness 
minimum ($\varphi=0.25-0.55$) was subtracted. Thus, the spectra shown in Fig.~\ref{harms} contain the radiation from the accretion spot only. The change in the hump positions is noticeable. 
This behaviour is also consistent with the assumption of the cyclotron 
origin of the humps, since the cyclotron harmonics should shift towards the red region as the angle 
$\theta$ between the magnetic field lines and the line of sight increases (see, for example, 
\cite{barrett85}, \cite{campbell08}). A spectacular feature in the presented spectra is a 
quasi-absorption near $\lambda=6075$~\AA. This feature does not change its position in the range 
of $\Delta \lambda=10$~\AA~and is absent in the out-of-eclipse brightness minimum of BS~Tri ($\varphi = 0.25-0.55$).

The position of cyclotron harmonics is determined mainly by the magnetic field strength $B$ of the rediated region. It is also influenced by electron temperature $T_e$, optical depth of emitting region and magnetic lines orientation defined by angle $\theta$. The same parameters affect the width of the cyclotron harmonics.  Thus, we can estimate the characteristics of the accretion spot by modeling cyclotron spectra. In the analysis of cyclotron 
spectra of polars, a homogeneous model of the 
cyclotron radiation zone \citep{barrett85, campbell08, kolbin19} has become widespread. 
It assumes that the radiating region is uniform in temperature and density, 
and the strength and direction of the magnetic field do not change significantly within the region. 
According to~\cite{ramaty69}, under the conditions of rapid Faraday rotation, the transfer equations 
for polarized radiation are simplified and reduced to independent equations for the ordinary (+) 
and extraordinary (--) waves. In the case of the assumption of homogeneity
of  the emitting region we have accepted, 
the solution of the transfer equation for both polarization modes has the form of 
\begin{equation}
I_{\pm} = \frac{ \mathfrak B}{2} (1-e^{-\alpha_{\pm} \Lambda}),
\label{rt_solution}
\end{equation}
where $\mathfrak B/2$ is the the Planck function for the polarization mode, 
$\alpha_{\pm}$ are the absorption 
coefficients, expressed in the units of $\omega^2_p/\omega_c c$ 
($\omega_c = e B/m_e c$ is the cyclotron frequency; 
$\omega_p = (4 \pi N_e e^2/m_e)^{1/2}$ --- plasma frequency, $N_e$ --- electron density). 
The dimensionless plasma parameter $\Lambda$ (size parameter) is determined as
\begin{equation}
\Lambda = \frac{\omega^2_p \ell}{\omega_c c}, 
\end{equation}
where $\ell$  is the depth of the radiating region along the line of sight. 
The total radiation intensity $I$ is found by summing the intensities of 
ordinary and extraordinary waves:
\begin{equation}
I = I_+ + I_-.
\label{total_int}
\end{equation}

Determination of the absorption coefficients $\alpha_{\pm}$ was carried out by means of convolution of the 
emission coefficients of single electrons with the relativistic Maxwell distribution and 
using the assumption of thermodynamic equilibrium of the medium where the Kirchhoff's law 
is satisfied (\cite{chan81, vaeth95}). To increase the computational speed of theoretical spectra, 
we calculated grids of absorption coefficients for a wide set of temperatures $T_e$, directions $\theta$ 
and frequencies $\omega/\omega_c$, which were then interpolated. 
The magnetic field strength and temperature were determined by the least squares method, 
which implies minimizing the function
\begin{equation}
\mathfrak L(B,T_e) = \sum_{i=1}^n \chi^2_i(B, T_e, \hat \theta_i, \hat \Lambda_i),
\end{equation}
where $n$ is the number of cyclotron spectra, $\chi^2_i$ is the sum of squared resudals between observed and theoretical fluxes in i$th$ spectrum, $\hat \theta_i$ and $\hat \Lambda_i$ are $\theta$ angle and $\Lambda$ parameters for the i$th$ observation respectively. The last parameters are determined as
\begin{equation}
(\hat \theta_i, \hat \Lambda_i) = \argmin_{\theta, \Lambda} \chi^2_i (B, T_e, \theta, \Lambda).
\end{equation}
In other words, the magnetic field strength and temperature in the accretion spot are searched while fitting the entire set of cyclotron spectra and the angle $\theta$ and size parameter $\Lambda$ are determined individually for each observed phase. 

Minimization of $\chi^2_i$ was carried out by the 
Gauss--Newton method. The optimization procedure was called many times for different initial approximations 
generated by the Monte-Carlo method, which guaranteed finding the global minimum of $\mathfrak L$. When estimating the parameters of the accretion spot, we used only the spectra ``a-b'', ``f-h'' (see Fig. \ref{harms}). We excluded the low-intensity spectra ``c-d'' from consideration due to errors in determination of the non-cyclotron background and low amplitude of the harmonics. We also excluded the dip in the range $6350-6800$~\AA~from the spectrum ``e'' which cannot be fitted by the used model of cyclotron radiation source. It is possible that the formation of this dip is associated with the self-absorption of cyclotron radiation at certain orientations of the accretion spot with respect to the observer. It should be noted that traces of this dip are also noticeable in the spectra ``a'', ``b'', ``f'', but its wavelength range is covered by the cyclotron harmonic. We also removed absorption feature at $6075$~\AA from the spectra before modeling as well as the region around 7000~\AA where we expect the Zeeman feature to appear (see below). 

The results of modeling the spectra with the most pronounced cyclotron harmonics are given in Table~\ref{tab_cyc}, 
and the comparison of the theoretical spectra with the observed ones is presented in Fig.~\ref{harms}. 
The magnetic field strength in the spot is found to be $B = 22.7 \pm 0.4$~MG, and the temperature $T_e \approx 7.5$~keV, 
which is lower than the temperature near the shock front 
$T_e \approx (3/8) \mu m_H G M_1 /k_B R_1 = 19 \pm 2$~keV by about a half ($\mu=0.5$ is the mean molecular weight of the hydrogen plasma). 
The angle between the magnetic field lines and the line of sight is close to $90^{\circ}$ for the most of spectra, which agrees with the 
assumption that the spot is near the edge of the stellar disk at the moments of maximum brightness. 
The size parameter $\Lambda \approx 10^7$ is typical for the regions of cyclotron emission in polars.

\begin{table}
\caption{Parameters of the cyclotron radiation region 
providing the best fit of
the observed spectra. 
The indices of the spectra corresponding to those shown in Fig.~\ref{harms}, obtained magnetic field 
strengths $B$, temperatures $T_e$, angles between the magnetic field lines and the line of sight $\theta$, 
as well as the logarithms of the  $\Lambda$ parameter are indicated.}
\label{tab_cyc}
\begin{center}
\begin{tabular}{cccccc}
\hline
ID 	&$\varphi$& $B$, MG 	& $T_e$, keV	& $\theta^{\circ}$ 	& $\log \Lambda$	\\ \hline
a	&0.642	& 22.7		& 7.5 		& 87.5  			& 7.3  \\
b	&0.705	& 22.7  	& 7.5 		& 86.6 				& 7.3  \\
e	&0.126	& 22.7		& 7.5		& 74.7 				& 7.4  \\
f	&0.189	& 22.7		& 7.5 		& 86.5 				& 7.4  \\
g	&0.633	& 22.7		& 7.5 		& 90.0 				& 7.1  \\
h	&0.633	& 22.7		& 7.5		& 89.1 				& 7.1  \\
\hline
\end{tabular}
\end{center}
\end{table}

\begin{figure}
\center
\includegraphics[scale=0.55]{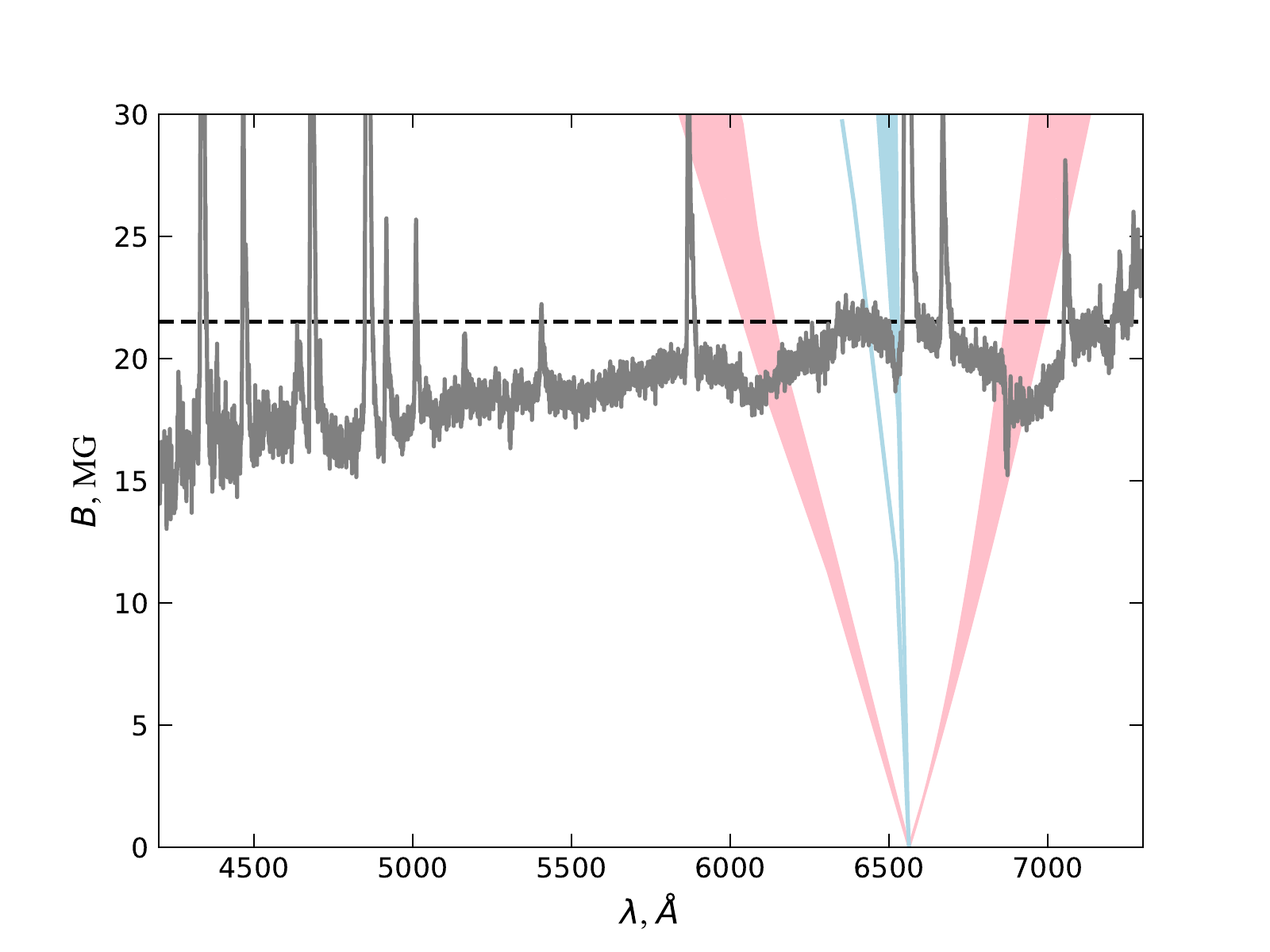}
\caption{Diagram of dependence  of the H$\alpha$ Zeeman splitting component positions from the magnetic field. 
The red area marks the position of the unresolvable $\sigma^+$ and $\sigma^-$ components, 
and the blue area --- of the $\pi$ components.  The diagram is superimpozed with the BS~Tri spectrum, 
which has an absorption feature near the H$\alpha$ line.}
\label{zeeman_feature}
\end{figure}

The `b' spectrum  has an absorption feature near the H$\alpha$ line at the wavelength of $\lambda=6522.4$~\AA. Fig~\ref{zeeman_feature} shows this spectrum plotted together with the dependance of position of H$\alpha$ line components varied with the magnetic field strength due to Zeeman effect 
borrowed from \citep{schmidt03}. It is seen that this feature can be interpreted as the $\pi$-component of the 
H$\alpha$ line, shifted to the blue region by the quadratic Zeeman effect in a magnetic field 
with an intensity $B=21.5 \pm 1$~MG. 
Moreover, for a given magnetic field, the position of the $\sigma^-$ component is in good agreement 
with a quasi-absorption at $6075$~\AA. An additional argument in favor of the ``Zeeman'' nature of this feature 
is the absence of noticeable (within $10$~\AA) changes in its position. Note that the dip near 
$\lambda=6920$~\AA~may be partially due to the $\sigma^+$ component of H$\alpha$, but the presence of 
telluric lines in this region prevents this assumption from being supported.

The Zeeman components of H$\alpha$  obviously do not get formed in the white dwarf atmosphere. 
In that case, their intensity would be at maximum in the phases of minimum brightness, which is not 
observed in the BS~Tri spectra. 
BS~Tri might embody the situation that underwent \citet{wickra87} when studying the spectra
of V834~Cen, and \cite{schwope97b} in the case of EP~Dra. They suggested that absorption 
features form in the cool halo surrounding the accretion spot.

\section{Light curve modeling}
\label{parlcmod}

As noted in Section \ref{morph}, in the lowered BS~Tri state, the dominant source of the 
out-of-eclipse brightness variability is the accretion spot. In this case, the variability of 
the radiation flux is due to a change in the area of the  spot projection on the picture plane, 
the passage of the spot behind the visible disk of the white dwarf, as well as the dependence 
of the intensity of cyclotron radiation on the direction of magnetic lines in relation to the observer. 
Thus, the behavior of the brightness of the polar is determined by the coordinates of the accretion spot 
and the orientation of the magnetic field lines in the spot. This makes it possible to find the position 
of the spot on the stellar surface, as well as the orientation of the magnetic dipole by modeling the light curves.

\begin{figure}
\center
\includegraphics[scale=0.9]{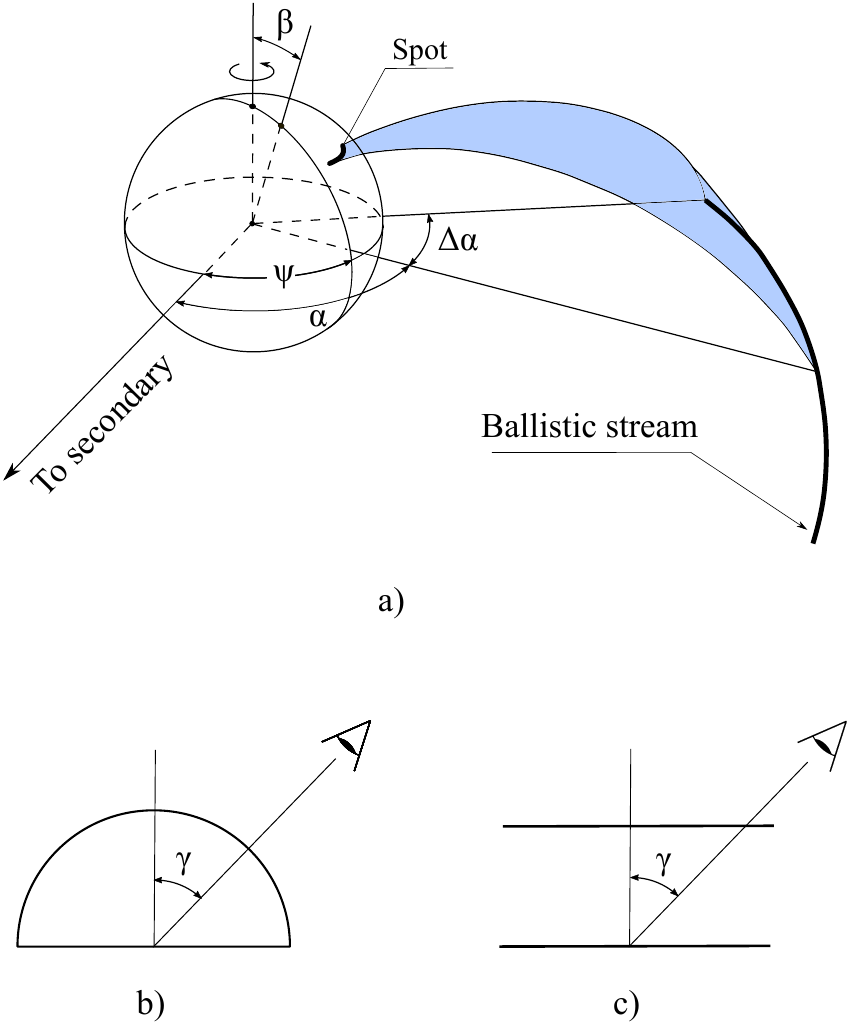}
\caption{a) Model of an accreting white dwarf in a polar. 
b) Hemispherical model of the radiation source. c) Plane-parallel model of the radiation source.}
\label{scheme_wd}
\end{figure}

To describe the light curves of BS~Tri, we used a simple model of an accreting white dwarf  with a dipole 
magnetic field (see Fig.~\ref{scheme_wd}a). The orientation of a magnetic dipole is determined by the 
inclination of its axis relative to the axis of rotation 
$\beta$ ($0^{\circ} \le \beta \le 180^{\circ}$), as well as by the longitude of the magnetic pole 
$\psi$ ($0^{\circ} \le \psi \le 360^{\circ}$), which is measured from the direction to the secondary 
in the direction of its orbital motion. The accretion spot is assumed to be geometrically thin and is 
``traced'' on the stellar surface by magnetic lines crossing the ballistic trajectory of the stream between 
the azimuthal angles $\alpha$ ($0^{\circ} \le \alpha \le 360^{\circ}$)  and $\alpha + \Delta \alpha$ ($\Delta \alpha \ge 0$).
Therefore, the segment of the ballistic trajectory lying between $\alpha$  and $\alpha + \Delta \alpha$,
imitates the stagnation region, where the ionized matter of the stream goes over to the magnetic trajectory 
and subsequently hits the surface of the white dwarf with the formation of an accretion spot. 
The calculation of the ballistic trajectory was carried out by solving the restricted three-body problem 
for particles escaping from the L$_1$ Lagrange point with a small momentum (see, for example, \cite{flannery75}). 
The integration of the equations of motion of the particles was carried out by the 
Runge--Kutta 4th order method.

The accretion spot was assumed to be uniform in density and temperature. It was also divided into small segments, 
and finding the radiation flux from it was reduced to integrating the intensity of cyclotron radiation 
over the observed part of the spot (which did not go beyond the disk), taking into account the orientation 
of the magnetic lines with respect to the observer. The fluxes in the photometric band were calculated 
by convolution of the spectral energy distribution with the filter transmission function. Cyclotron spectra 
were calculated using the  (\ref{rt_solution})--(\ref{total_int}) formulas. We refer the reader to the study of \cite{kolbin20} for details of the mapping algorithm used here.

We used two models of the emitting region. 
The first (hereinafter, the $\Lambda$-const model) is quite often used to analyze photometric and polarization 
observations of polars and assumes independence of the $\Lambda$ parameter from the rotational phase. 
Therefore, the spot can be represented as consisting of a set of emitting hemispheres, for which the 
size along the line of sight $\ell$ does not depend on the angle of aspect from which the observations 
are made (see Fig.~\ref{scheme_wd}~b). In the second model $\ell \sim 1/ \cos \gamma$, where $\gamma$ 
is the angle between the surface normal and the line of sight (see Fig.~\ref{scheme_wd}c). 
In this case, the spot can be represented as a thin plane-parallel layer.

\begin{figure}
\center
\includegraphics[width=\columnwidth]{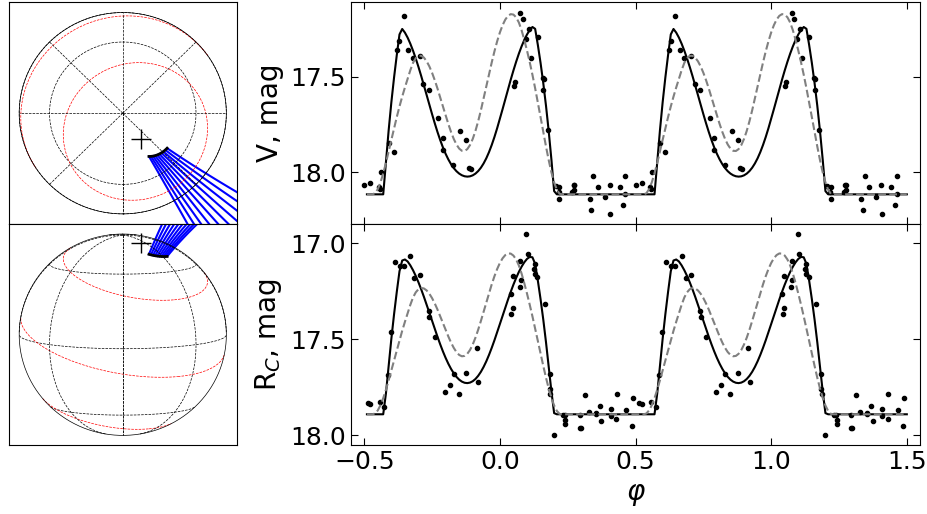}
\caption{Modeling results for BS~Tri light curves. Left: the found model of a white dwarf, 
observed from the side of the rotation pole (top panel) and from the observer 
at $\varphi=0$ (bottom panel). This model was found in the approximation of plane-parallel 
radiation sources. The cross shows the position of the magnetic pole, the red lines describe 
the magnetic parallels, the blue lines --- the magnetic dipole lines emerging from the spot. 
Right: a comparison of the observed polar light curves (dots) and simulated light curves 
obtained within the $\Lambda$-const model (the dashed line) and the plane-parallel source 
model (the solid line). The top panel shows a comparison of the V-band light curves, and 
the bottom one shows the R$_C$ light curves.}
\label{lc_mod}
\end{figure}

We have modeled the BS~Tri light curves in the Cousins system $V$ and R$_C$ bands (see Fig. \ref{lc_mod}). 
The light curves were obtained with the Zeiss-1000 telescope of the SAO RAS in the nights 
from August 29 to 31, 2019. They have a flat plateau around $\varphi=0.5$ and a 
pronounced two-humped maximum, which indicates a lowered state of the polar during the 
observation period. The contribution of white and red dwarfs was subtracted from the light 
curves, which was taken to be equal to the mean brightness at the plateau.

The observed light curves were fitted by the least squares method with a 
search for the orientation of the magnetic dipole, i.e. the  $\beta$ and $\psi$ parameters, 
as well as the  $\alpha$  and $\Delta \alpha$ angles, which determine the position of the 
stagnation region. Due to the nonlinearity of the inverse problem, the minimized sum of  
squared residuals can have many local minima.

For this reason, the search for an approximate solution was carried out using a genetic algorithm 
(see, e.g., \cite{charbon95}). This method reliably found the vicinity of the global minimum; however, 
the further convergence to the exact solution turned out to be rather slow. 
Therefore, the second stage was introduced into the minimization procedure for $\chi^2$, 
where the exact solution is found using the quasi-Newton 
Broyden--Fletcher--Goldfarb--Shanno algorithm (see, for example, \cite{fletcher87}).

The light curve shapes depend on the $\Lambda$-parameter of the emitting region. 
Unfortunately, we cannot determine this parameter in an independent way. The  
$\Lambda$-parameter estimates found in the previous section can differ greatly from 
its true value during the period of photometric observations. Therefore, the search 
for the $\Lambda$-parameter was carried out by describing the light curves of the polar 
simultaneously with the $\beta$, $\psi$, $\alpha$, $\Delta \alpha$ parameters. 
The temperature of the accretion spot was fixed at $T_e = 10$~keV, 
which is consistent with the results of  cyclotron spectra modeling.

The results of modeling the BS~Tri light curves are shown in Fig.~\ref{lc_mod}. 
We can see that the $\Lambda$-const model is unable to satisfactorily describe the 
sharp emission peaks that appear when the spot is close to the edge of the stellar disk. 
In the approximation of plane-parallel sources,  increasing radiation intensity 
as the spot approaches the edge of the disk occurs not only owing to an increase in the 
angle $\theta$ between the magnetic field lines and the direction to the observer, 
but also due to an increase in the optical depth along the line of sight. 
The combination of both of these effects leads to a satisfactory description of the observed data. 
The polar parameters obtained within the framework of the model of plane-parallel sources 
are listed in Table~\ref{tab_lcmod}. Note that in view of a poor description of 
observations by the $\Lambda$-const model, its results are not suitable for arguing about
the geometric parameters of accretion in the polar.

\begin{table}
\caption{The results of fitting of the BS~Tri light curves by the model of plane-parallel sources. 
The shown $\Lambda$-parameter corresponds to the layer thickness along the normal to the surface.}
\label{tab_lcmod}
\begin{center}
\begin{tabular}{ll}
\hline
Parameter 				& Value		\\ \hline
$\beta^{\circ}$			& $18 \pm 4$		  \\
$\psi^{\circ}$			& $34 \pm 7$ 		  \\
$\alpha^{\circ}$		& $26 \pm 10$		  \\
$\Delta \alpha^{\circ}$	& $30 \pm 10$ 		  \\
$\log \Lambda$				&  $(3.5 \pm 0.5)$		  \\
\hline
\end{tabular}
\end{center}
\end{table}

\section{Doppler tomography}
\label{pardoptom}

The essence of the Doppler tomography method is to search for the distribution of 
emission regions in a two-dimensional velocity space with a description of the behavior 
of a spectral line during the orbital period. Each point in the velocity space can be 
determined by the polar coordinates $v$ and $\vartheta$. Here, $v$ is the modulus of the 
emitting point velocity (relative to the center of mass of the system) multiplied by 
$\sin i$, and  $\vartheta$ is the angle between the velocity vector of the emitting point 
and the line connecting the centers of mass of the system components. The $\vartheta$ 
angle is reckoned in the direction of motion of the secondary. 
We refer the reader to the studies of \cite{kotze15, marsh16} for the details of 
Doppler tomogram interpretation.

The BS~Tri tomography was carried out by the maximum entropy method implemented in the 
doptomog software package by \cite{kotze15}. The programs within the package allow
building Doppler maps both in a standard projection, where the velocity $v$ increases 
from the center to the periphery of the map, and in an inside-out projection, where the 
velocity $v$ increases from the periphery to the center. The first method is convenient 
for studying low-velocity structures, for example, an irradiated hemisphere of the secondary component 
or a part of an accretion stream near the Lagrange point L$_1$. Around the white dwarf, 
the gas accelerates to high velocities and its trace on the Doppler map in the standard 
projection is smeared over a large area. An inside-out projection is preferable
for the study of such high-velocity structures, in which they will be concentrated in a more compact 
area and have a greater contrast. In addition, we used a flux-modulated tomography version, 
which assumes a sinusoidal variation in the intensity of the emitting points during the 
orbital period \cite{steeghs03}. This option is more preferable for studying the optically 
thick media. Before performing the Doppler mapping, the continuum was subtracted from the spectra, 
approximated by an algebraic polynomial with the removal of features that strongly deviate from it. 
The mapping was performed based on the HeII~$\lambda$4686 line profiles, which have a 
pronounced two-peak structure and a rather high signal-to-noise ratio. 
We studied the spectra of two sets of observations (2011 and 2012), covering the orbital 
period of the polar. The spectra falling on the eclipse phases were excluded from consideration.

\begin{figure*}
\center
\includegraphics[scale=0.5]{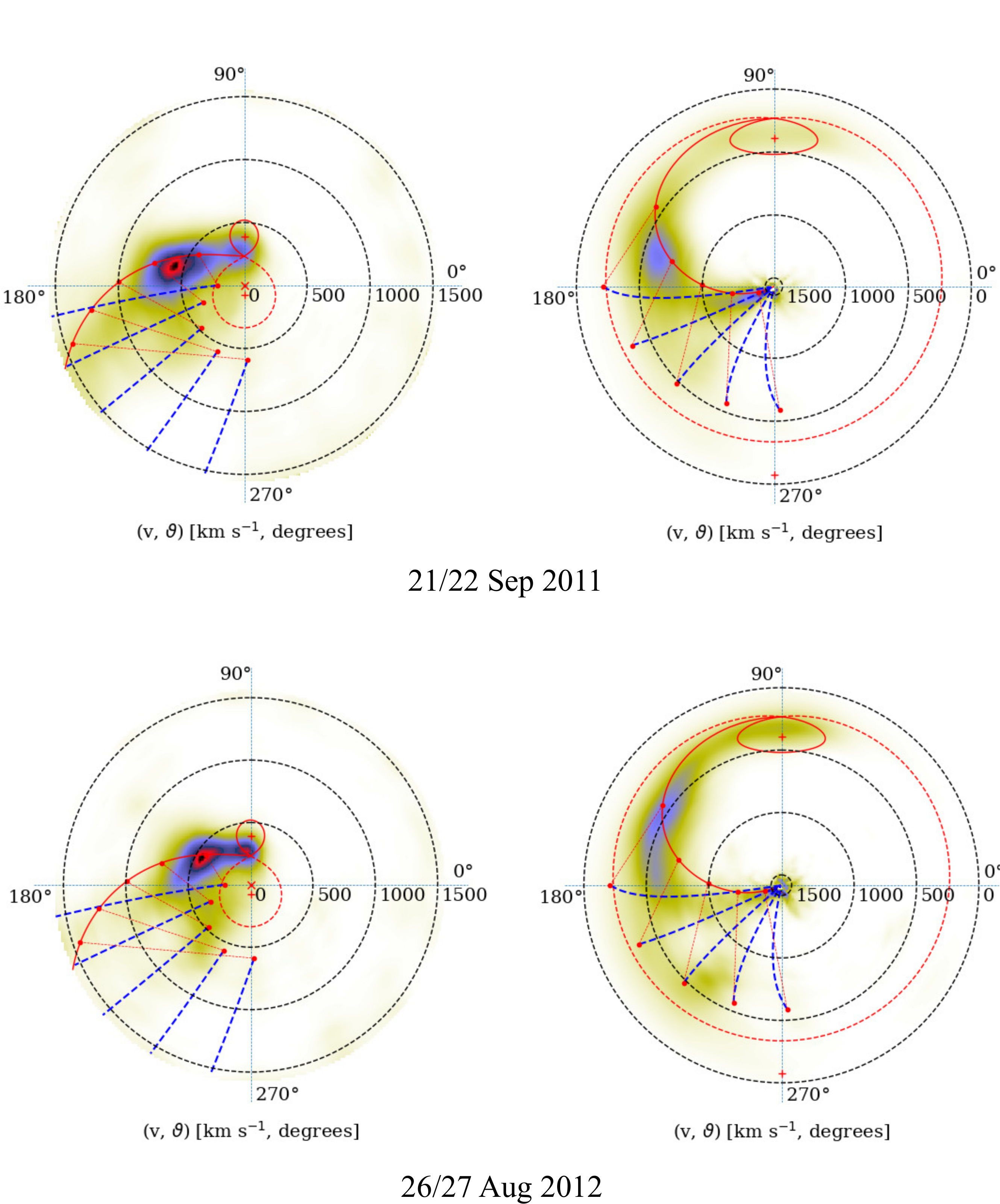}
\caption{Doppler tomograms of BS~Tri for HeII~$\lambda$4686 line according to the 2011 (top) and 2012 (bottom)  
spectroscopic observations. 
The left plot presents the tomograms in the standard projection, while the right plot---in the inside-out one. 
The velocities of the Roche lobes of the primary (the red dashed closed line) and secondary (the red continuous closed line) 
components are superimposed on the tomograms. In addition, the velocities of the ballistic trajectory
(the red continuous line) and the magnetic part of the trajectory (blue dashed lines) are superimposed on the tomograms.}
\label{dopptom2}
\end{figure*}

The BS~Tri tomograms based on the data from two sets of observations in standard and inside-out projections 
are shown in Fig.~\ref{dopptom2}. 
The Roche lobe velocities for both componnets were superimposed on the Doppler maps. 
In addition, the maps were overplotted by the particle velocities on the ballistic trajectory and on the magnetic trajectories approximated by the dipole. The calculations were carried out based on the system parameters found above ($M_1 = 0.6 M_{\odot}$, $Q = 5.12$, $i=85^{\circ}$). 
 The magnetic dipole was oriented 
according to the results of the light curve modeling performed in the previous section 
(see Table~\ref{tab_lcmod}). The figure shows the velocities of particles on magnetic lines intersecting with the ballistic trajectory at azimuthal angles $\alpha = 10-70^{\circ}$. 

Standard projection tomograms show two bright areas. 
The first one is located on the surface of the red dwarf, facing towards the primary component. 
The second bright area is in the accretion stream near the expected trace of the ballistic part. 
Note that the maps obtained according to the 2011 data are more blurred due to the lower 
spectral resolution. We can see that the position of the stream near the Lagrange point L$_1$ 
differs according to the data of 2011 and 2012. This confirms our assumption about the 
existence of certain effects that are not taken into account in the restricted three-body problem. 
The position of the radiation maximum of the stream component also changes. 
During the 2011 observation period, the brightness maximum moved at a higher velocity 
($v \approx 600$~km/s), compared to the  2012 observation period of 
($v \approx 450$~km/s). In addition, a larger $\vartheta $ angle is notable in 2011. 
Probably, a deeper penetration of the stream into the magnetosphere of the white dwarf was observed in 2011. 
The 2012 tomograms of the  polar demonstrate the emission region near 
$v\approx 500$~km/s  and $\vartheta \approx 135^{\circ}$.
A similar structure also appears on the 2011 tomograms, but it is more blurred. 
The model particle velocities on the magnetic dipole lines indicate that this structure is formed on the magnetic part of the trajectory near the orbital plane. 
We can see that the transition to a magnetic trajectory occurs in the range of azimuthal 
angles  $\alpha = 15 - 55^{\circ}$. Note that it is not possible to accurately determine the 
lower boundary for $\alpha$ due to the poor separation of the magnetic and ballistic 
trajectories on the tomograms in standard and inside-out projections. The indicated $\alpha$ 
range is in good agreement with the position of the stagnation region found by modeling 
the light curves (see Table~\ref{tab_lcmod}).

\section{Conclusion}

We have analyzed here the spectra and light curves of the BS~Tri eclipsing polar. 
Analysis of the photometric data shows variations in the shape of the light curves 
caused by different contributions of the accretion stream to the integral radiation of the system. 
Based on the radial velocity curves of the irradiated hemisphere of the secondary component, 
the orbital parameters of the system and its components have been refined: 
$Q = 5.12 \pm 0.35$, $M_1 = 0.60 \pm 0.04 M_{\odot}$, $i=85\pm0.5^{\circ}$.
The masses of the components we found differ from those presented in (\cite{borisov15}), 
where the radial velocities of the irradiated surface of secondary are approximated at a fixed velocity 
of the center of mass $\gamma$, roughly estimated from the radial velocities 
of  lines without separating them into components. We show that the theoretical 
velocities of the accretion stream near the Lagrange point L$_1$ do not agree with the 
Doppler tomograms of BS~Tri. A possible explanation for this mismatch is the 
contribution of the red dwarf's magnetic fields to the stream flowing out of it. 
Difficulties in describing the stream velocities near the L$_1$ Lagrange point are also 
noted for the polar AM~Her (\cite{schwarz02, staude04}). The BS~Tri spectra exhibit 
broad cyclotron harmonics that change their position as the white dwarf spins. 
By modeling the cyclotron spectra, the magnetic field strength in the accretion spot 
($B = 22.7 \pm 0.4$~MG), as well as the average spot temperature ($T_e\sim 10$~keV) have been found. 

An interesting feature of BS~Tri is the presence in its spectra of absorption components 
of the Zeeman splitting of the H$\alpha$ line, which appear simultaneously with cyclotron 
harmonics and are formed at  magnetic field strength of  $B=21.5 \pm 1$~MG. 
A possible explanation for this phenomenon is the presence of a cool halo around the acreation spot, 
whose  signs are found in some other polars (\citet{wickra87, schwope97b}). 
The orientation of the magnetic dipole and the position of the stagnation region have been estimated 
by modeling the light curves of the polar in two photometric bands. 
We show that the shape of the BS~Tri light curves cannot be explained by the variability 
of the accretion spot radiation intensity, associated only with a change in the orientation 
of the magnetic field lines with respect to the observer. A satisfactory description of the 
observations is achieved under the assumption of a plane-parallel spot structure, 
where the optical depth is proportional to $1/\cos \gamma$. Doppler tomograms of BS~Tri 
in the HeII~$\lambda$4686 line demonstrate the change in the position of the emission source
between two sets of spectroscopic observations. The constraint on the position of the 
stagnation region ($\alpha < 55^{\circ}$) found from the Doppler maps is in a good 
agreement with the results of the polar's light curve modeling.

\section*{Acknowledgements}
The study was carried out with the financial support of the Russian Foundation for Basic Research 
within the framework of scientific project No.~19-32-60048. The work of N. Katysheva and S. Shugarov was supported by the Program of Development of Lomonosov Moscow State University ``Leading Scientific and  Educational Schools'', project ``Fundamental and Applied Space Research''. The work by S. Shugarov was also supported by the Slovak Research and Development Agency under the  contract No. APVV-15-0458 and by the Slovak Academy of Sciences grant VEGA No. 2/0008/17. We deeply appreciate the significant contribution of the referee to the improvement of the paper. 

\section*{Data Availability}
The data underlying this article will be shared on reasonable re-quest to the corresponding author.

\renewcommand{\refname}{REFERENCES}

\bsp	
\label{lastpage}
\end{document}